\documentclass[10pt]{article}
\usepackage{graphicx}
\usepackage{amsmath}
\usepackage{amssymb}
\usepackage{caption2}
\begin{document}
\begin{center}
{\large {\bf \sc{  Analysis of the axialvector doubly heavy tetraquark states  with QCD sum rules
  }}} \\[2mm]
Zhi-Gang  Wang \footnote{E-mail: zgwang@aliyun.com.  }   \\
 Department of Physics, North China Electric Power University, Baoding 071003, P. R. China
\end{center}

\begin{abstract}
In this article, we construct the axialvector-diquark-scalar-antidiquark type currents to interpolate the axialvector doubly heavy tetraquark states, and study them with the QCD sum rules in details by carrying out the operator product expansion up to the vacuum condensates of dimension 10.
\end{abstract}



\section{Introduction}
The scattering amplitude for one-gluon exchange  is proportional to
\begin{eqnarray}
t^a_{jm}t^a_{kn}&=&-\frac{2}{3}\frac{\delta_{jk}\delta_{mn}-\delta_{jn}\delta_{km}}{2}
 +\frac{1}{3}\frac{\delta_{jk}\delta_{mn}+\delta_{jn}\delta_{km}}{2} \nonumber\\
 &=&-\frac{1}{3}\varepsilon_{ijm}\,\varepsilon_{ikn} +\frac{1}{6}S_{jm,kn} \, ,
\end{eqnarray}
where $t^a=\frac{\lambda^a}{2}$, the $\lambda^a$ is the  Gell-Mann matrix, $S_{jm,kn}=\delta_{jk}\delta_{mn}+\delta_{jn}\delta_{km}$, the $i$, $j$, $k$, $m$ and $n$ are color indexes.  The negative sign in front of the antisymmetric  antitriplet $\bar{3}$ indicates the interaction
is attractive  while the positive sign in front of the symmetric sextet $6$ indicates
 the interaction  is repulsive, the attractive interaction  favors  formation of
the diquarks in  color antitriplet while the repulsive  interaction  disfavors  formation of
the diquarks in  color sextet \cite{One-gluon}.   We can construct both the $\bar{3}\otimes 3$ type currents $\varepsilon_{ijk}\varepsilon_{imn}Q^T_jC\Gamma Q_k \bar{q}_m\Gamma^\prime C\bar{q}^{\prime T}_n$ and the $6\otimes \bar{6}$ type currents $S_{jk,mn}Q^T_jC\Gamma Q_k \bar{q}_m\Gamma^\prime C\bar{q}^{\prime T}_n$ satisfying Fermi-Dirac statistics to interpolating the doubly heavy  tetraquark states, where the $\Gamma$ and $\Gamma^\prime$ are the Dirac $\gamma$ matrixes. If there really exist the $6\otimes \bar{6}$ type doubly charmed tetraquark states, they should have      much larger masses than the corresponding $\bar{3}\otimes 3$ type tetraquark states with the same quantum numbers.
The color antitriplet  diquarks $\varepsilon^{ijk} Q^{T}_j C\Gamma Q_k$ with $Q=c$ or $b$
only  have  two  structures  in Dirac spinor space, where $\Gamma=\gamma_\mu $ and $\sigma_{\mu\nu}$ for the axialvector  and  tensor diquarks, respectively.
The axialvector diquarks $\varepsilon^{ijk} Q^{T}_j C\gamma_{\mu} Q_k$ are more stable than the tensor diquarks $\varepsilon^{ijk} Q^{T}_j C\sigma_{\mu\nu} Q_k$, it is better to choose the axialvector diquarks $\varepsilon^{ijk} Q^{T}_j C\gamma_{\mu} Q_k$ to construct the ground state doubly heavy tetraquark states.

In 2016, the LHCb collaboration observed the doubly charmed baryon state  $\Xi_{cc}^{++}$ in the $\Lambda_c^+ K^- \pi^+\pi^+$ mass spectrum in
a $pp$ data sample collected by LHCb at $\sqrt{s} = 13\,\rm{TeV}$ with a signal yield of $313\pm33$, and measured the mass, but did not determine the spin   \cite{LHCb-Xicc}. The $\Xi_{cc}^{++}$ maybe have the spin $\frac{1}{2}$ or $\frac{3}{2}$, we can take the diquark $\varepsilon^{ijk}  c^T_iC\gamma_\mu c_j$ as basic constituent to construct the current
\begin{eqnarray}
J_{\Xi_{cc}}(x)&=& \varepsilon^{ijk}  c^T_i(x)C\gamma_\mu c_j(x)
\gamma_5\gamma^\mu u_k(x)  \, ,
\end{eqnarray}
or
\begin{eqnarray}
J_{\Xi_{cc}}^\mu(x)&=& \varepsilon^{ijk}  c^T_i(x)C\gamma^\mu c_j(x) u_k(x)  \, ,
\end{eqnarray}
to study it with the QCD sum rules \cite{Xicc-QCDSR}.

Up to now, no experimental candidates for the tetraquark configurations $QQ\bar{q}\bar{q}^\prime$ or $qq^{\prime}\bar{Q}\bar{Q}$ have been observed.
The observation of the doubly charmed baryon state  $\Xi_{cc}^{++}$ has led a renaissance in the doubly heavy tetraquark spectroscopy.
In this article, we choose the axialvector diquarks $\varepsilon^{ijk} Q^{T}_j C\gamma_{\mu} Q_k$ to construct the currents to interpolate the doubly heavy tetraquark states.
 There have been many  works on the doubly heavy tetraquark states, such as potential quark models \cite{JMRichard-1982,Exist-Stable-poten-et-al,QQ-quark-model,Color-Magnetic,KR-PRL} or constituent  diquark models \cite{Polosa-diquark-model}, QCD sum rules \cite{Nielsen-Lee,QQ-QCDSR,QQ-QCDSR-Chen}, heavy quark symmetry \cite{QQ-heavy-quark-Manohar,QQ-heavy-quark-Karliner,QQ-HQS,EQ-PRL}, lattice QCD \cite{QQ-Latt,QQ-Latt-mass1,QQ-Latt-mass2}, etc.
If  the two heavy quarks are in a long
separation, the gluon exchange force between them is screened by the two light quarks, then a loosely $Q\bar{q}-Q\bar{q}^\prime$ type
bound state is formed. On the other hand, if the two heavy quarks are in a short separation, the heavy
$QQ$ pair forms a compact point-like color source in heavy quark limit, and attracts a  light $\bar{q}\bar{q}^\prime$ pair which serves as another compact point-like color source,   then an exotic $QQ-\bar{q}\bar{q}^\prime$ type tetraquark state  is formed.
The existence and stability of the $QQ\bar{q}\bar{q}^\prime$ tetraquark states have been extensively discussed in early literatures based on the potential models \cite{JMRichard-1982,Exist-Stable-poten-et-al} and heavy quark symmetry \cite{QQ-heavy-quark-Manohar}, while the existing  doubly heavy tetraquark mass spectra  differ from each other in one way or the other  \cite{QQ-quark-model,Color-Magnetic,KR-PRL,Polosa-diquark-model,Nielsen-Lee,QQ-QCDSR,QQ-QCDSR-Chen,QQ-heavy-quark-Karliner,QQ-HQS,EQ-PRL,QQ-Latt-mass1,QQ-Latt-mass2}.  More theoretical and experimental works are still needed.

The  QCD sum rules is a powerful nonperturbative theoretical tool in studying the
ground state hadrons, and has given many successful descriptions of the hadronic properties \cite{SVZ79,Reinders85,NarisonBook}.
Although the  doubly heavy tetraquark states have been studied with the QCD sum rules, the energy scale dependence of the QCD sum rules has  not been studied yet. In Refs.\cite{WangHuang-3900,Wang-4660-2014,Wang-4025-CTP,WangHuang-NPA-2014,WangHuang-mole}, we observe that in the QCD sum rules for the hidden-charm (or hidden-bottom) tetraquark states and molecular states, the integrals
 \begin{eqnarray}
 \int_{4m_Q^2(\mu)}^{s_0} ds \rho_{QCD}(s,\mu)\exp\left(-\frac{s}{T^2} \right)\, ,
 \end{eqnarray}
are sensitive to the heavy quark masses $m_Q(\mu)$, where the $\rho_{QCD}(s,\mu)$ denotes the QCD spectral densities and the $T^2$ denotes the Borel parameters.
Variations of the heavy quark masses $m_Q(\mu)$ or the energy scales  $\mu$ lead to changes of integral ranges $4m_Q^2(\mu)-s_0$ of the variable  $ds$ besides the QCD spectral densities $\rho_{QCD}(s,\mu)$,
therefore changes of the Borel windows and predicted masses and pole residues.  In this article, we revisit the QCD sum rules for the axialvector doubly heavy tetraquark states and choose the optimal energy scales to extract the masses.

The article is arranged as follows:  we derive the QCD sum rules for the masses and pole residues of  the
axialvector  doubly heavy tetraquark states in Sect.2;  in Sect.3, we present the numerical results and discussions; and Sect.4 is reserved for our
conclusion.

\section{The QCD sum rules for  the  axialvector   doubly heavy tetraquark states }
In the following, we write down  the two-point correlation functions $\Pi^{J}_{\mu\nu}(p)$ and $\Pi^{\eta}_{\mu\nu}(p)$  in the QCD sum rules,
\begin{eqnarray}
\Pi_{\mu\nu}^{J/\eta}(p)&=&i\int d^4x e^{ip \cdot x} \langle0|T\left\{J/\eta_\mu(x) J/\eta_\nu^{\dagger}(0)\right\}|0\rangle \, ,
\end{eqnarray}
where
\begin{eqnarray}
J_\mu(x)&=&\varepsilon^{ijk}\varepsilon^{imn} \, Q^{T}_j(x)C\gamma_\mu Q_k(x) \,\bar{u}_m(x)\gamma_5C \bar{s}^T_n(x) \, , \\
\eta_\mu(x)&=&\varepsilon^{ijk}\varepsilon^{imn} \, Q^{T}_j(x)C\gamma_\mu Q_k(x) \,\bar{u}_m(x)\gamma_5C \bar{d}^T_n(x) \, ,
\end{eqnarray}
$Q=c,b$, the  $i$, $j$, $k$,  $m$, $n$ are color indexes, the $C$ is the charge conjugation matrix.
In the type-II diquark-antidiquark model \cite{Maiani-II-type}, the building blocks (diquark and antidiquark) are taken
  as point-like color sources,  the size of the entire tetraquark is consistently larger than the size of its building blocks,
    the spin-spin interactions between the quarks and antiquarks in the effective Hamiltonian in the type-I diquark-antidiquark model \cite{Maiani-3872} are neglected. The mass spectrum derived in the type-II diquark-antidiquark model is superior to that derived in the type-I diquark-antidiquark model, and is compatible with the experimental data.
The tetraquark states are spatial extended objects, not point-like objects, while  we choose the local currents to interpolate the tetraquark states in the QCD sum rules, and take all the quarks and antiquarks as the color sources, the finite size effects are neglected, which leads to some uncertainties.

On the phenomenological side,  we insert  a complete set of intermediate hadronic states with
the same quantum numbers as the current operators $J_\mu(x)$ and $\eta_\mu(x)$ into the
correlation functions $\Pi^{J}_{\mu\nu}(p)$ and $\Pi^{\eta}_{\mu\nu}(p)$  respectively to obtain the hadronic representation
\cite{SVZ79,Reinders85}, and isolate the ground state
contributions,
\begin{eqnarray}
\Pi^{J/\eta}_{\mu\nu}(p)&=&\frac{\lambda_{Z}^2}{M^2_{Z}-p^2} \left( -g_{\mu\nu}+\frac{p_\mu p_\nu}{p^2}\right) +\cdots  \nonumber\\
&=&\Pi_{J/\eta}(p^2)\left( -g_{\mu\nu}+\frac{p_\mu p_\nu}{p^2}\right) +\cdots \, ,
\end{eqnarray}
where the pole residues  $\lambda_{Z}$ are defined by $ \langle 0|J/\eta_\mu(0)|Z_{QQ}(p)\rangle=\lambda_{Z}\,\varepsilon_\mu$, the $\varepsilon_\mu$ are
the polarization vectors of the axialvector tetraquark states  $Z_{QQ}$.

The current $J_\mu(x)$ can be rewritten as
\begin{eqnarray}
J_\mu(x)&=& Q^{T}_j(x)C\gamma_\mu Q_k(x) \left[\bar{u}_j(x)\gamma_5C \bar{s}^T_k(x)-\bar{u}_k(x)\gamma_5C \bar{s}^T_j(x)\right]   \nonumber\\
&=& \frac{1}{2}\left[Q^{T}_j(x)C\gamma_\mu Q_k(x)-Q^{T}_k(x)C\gamma_\mu Q_j(x) \right]\left[\bar{u}_j(x)\gamma_5C \bar{s}^T_k(x)-\bar{u}_k(x)\gamma_5C \bar{s}^T_j(x)\right] \, ,
\end{eqnarray}
according to the identity  $\varepsilon_{ijk}\,\varepsilon_{imn}=\delta_{jm}\delta_{kn}-\delta_{jn}\delta_{km}$ in the color space. The current $J_\mu(x)$ is of $\bar{3}\otimes 3$ type  in both the color space and flavor space, we can also construct the current $\widetilde{J}_\mu(x)$ satisfying Fermi-Dirac statistics, which is of  $6\otimes \bar{6}$ type in the color space and $\bar{3}\otimes 3$ type  in the flavor space, and differs from the corresponding current constructed in Ref.\cite{QQ-QCDSR-Chen} slightly,
\begin{eqnarray}
\widetilde{J}_\mu(x)&=&\frac{1}{2} \left[Q^{T}_j(x)C\gamma_5 Q_k(x)+Q^{T}_k(x)C\gamma_5 Q_j(x)\right] \left[\bar{u}_j(x)\gamma_\mu C \bar{s}^T_k(x)+\bar{u}_k(x)\gamma_\mu C \bar{s}^T_j(x)\right] \, .
\end{eqnarray}
The attractive  interaction induced by one-gluon exchange favors formation of the color antitriplet diquark state $Q^{T}_j(x)C\gamma_\mu Q_k(x)-Q^{T}_k(x)C\gamma_\mu Q_j(x)$, while the repulsive  interaction induced by one-gluon exchange disfavors formation
of the color sextet  diquark state $Q^{T}_j(x)C\gamma_5 Q_k(x)+Q^{T}_k(x)C\gamma_5 Q_j(x)$. If there really exists a doubly charmed tetraquark state $\widetilde{Z}_{QQ}$, which couples potentially
to the current $\widetilde{J}_\mu(x)$, then the tetraquark state $\widetilde{Z}_{QQ}$ should have much larger mass than the corresponding tetraquark state $Z_{QQ}$.
As the color magnetic interaction $-\sum_{i<j} C_{ij}\lambda_{i}\cdot \lambda_{j}\, \sigma_{i}\cdot \sigma_{j}$ leads to mixing between the tetraquark states $Z_{QQ}$ and $\widetilde{Z}_{QQ}$, where the $\lambda_i$ and $\sigma_i$ denote the Gell-Mann
matrices and  Pauli matrices, respectively \cite{One-gluon,Color-Magnetic}.
Some $6\otimes \bar{6}$ type components  in the color space can lead to larger predicted tetraquark mass than the $M_{Z}$, for example, if we take the replacement,
\begin{eqnarray}
J_\mu(x)&\to& J_\mu(x)\,\cos\theta+\widetilde{J}_\mu(x)\,\sin\theta \, ,
\end{eqnarray}
we expect to obtain a tetraquark mass $M$ with the value $M_{Z}< M<M_{\widetilde{Z}}$. The conclusion survives for the current $\eta_\mu(x)$. However, in Ref.\cite{QQ-QCDSR-Chen}, M. L. Du et al obtain degenerate masses for the $Z_{QQ}$ and $\widetilde{Z}_{QQ}$ based on the QCD sum rules. This subject needs to be further studied.

 In the following,  we briefly outline  the operator product expansion for the correlation functions  $\Pi^{J}_{\mu\nu}(p)$ and $\Pi^{\eta}_{\mu\nu}(p)$ in perturbative QCD.  We contract the $u$, $d$, $s$  and $Q$ quark fields in the correlation functions  $\Pi^{J}_{\mu\nu}(p)$ and $\Pi^{\eta}_{\mu\nu}(p)$ with Wick theorem, and obtain the results:
 \begin{eqnarray}
 \Pi_{\mu\nu}^J(p)&=&-2i\varepsilon^{ijk}\varepsilon^{imn}\varepsilon^{i^{\prime}j^{\prime}k^{\prime}}\varepsilon^{i^{\prime}m^{\prime}n^{\prime}} \int d^4x e^{ip \cdot x}   \nonumber\\
&&{\rm Tr}\left[ \gamma_{\mu}S_Q^{kk^{\prime}}(x)\gamma_{\nu} CS_Q^{Tjj^{\prime}}(x)C\right] {\rm Tr}\left[ \gamma_{5}U^{m^{\prime}m}(-x)\gamma_{5} CS^{Tn^{\prime }n}(-x)C\right]     \, , \\
 \Pi_{\mu\nu}^\eta(p)&=&-2i\varepsilon^{ijk}\varepsilon^{imn}\varepsilon^{i^{\prime}j^{\prime}k^{\prime}}\varepsilon^{i^{\prime}m^{\prime}n^{\prime}} \int d^4x e^{ip \cdot x}   \nonumber\\
&&{\rm Tr}\left[ \gamma_{\mu}S_Q^{kk^{\prime}}(x)\gamma_{\nu} CS_Q^{Tjj^{\prime}}(x)C\right] {\rm Tr}\left[ \gamma_{5}U^{m^{\prime}m}(-x)\gamma_{5} CD^{Tn^{\prime }n}(-x)C\right]     \, ,
\end{eqnarray}
 where the $U^{ij}(x)$, $D^{ij}(x)$, $S^{ij}(x)$ and $S^{ij}_Q(x)$ are the full  $u$, $d$, $s$ and $Q$ quark propagators, respectively \cite{Reinders85,Pascual-1984},
\begin{eqnarray}
U/D_{ij}(x)&=& \frac{i\delta_{ij}\!\not\!{x}}{ 2\pi^2x^4}
 -\frac{\delta_{ij}\langle
\bar{q}q\rangle}{12}  -\frac{\delta_{ij}x^2\langle \bar{q}g_s\sigma Gq\rangle}{192} -\frac{ig_s G^{a}_{\alpha\beta}t^a_{ij}(\!\not\!{x}
\sigma^{\alpha\beta}+\sigma^{\alpha\beta} \!\not\!{x})}{32\pi^2x^2}\nonumber\\
&& -\frac{1}{8}\langle\bar{q}_j\sigma^{\mu\nu}q_i \rangle \sigma_{\mu\nu} +\cdots \, ,
\end{eqnarray}

\begin{eqnarray}
S_{ij}(x)&=& \frac{i\delta_{ij}\!\not\!{x}}{ 2\pi^2x^4}
-\frac{\delta_{ij}m_s}{4\pi^2x^2}-\frac{\delta_{ij}\langle
\bar{s}s\rangle}{12} +\frac{i\delta_{ij}\!\not\!{x}m_s
\langle\bar{s}s\rangle}{48}-\frac{\delta_{ij}x^2\langle \bar{s}g_s\sigma Gs\rangle}{192}+\frac{i\delta_{ij}x^2\!\not\!{x} m_s\langle \bar{s}g_s\sigma
 Gs\rangle }{1152}\nonumber\\
&& -\frac{ig_s G^{a}_{\alpha\beta}t^a_{ij}(\!\not\!{x}
\sigma^{\alpha\beta}+\sigma^{\alpha\beta} \!\not\!{x})}{32\pi^2x^2}  -\frac{1}{8}\langle\bar{s}_j\sigma^{\mu\nu}s_i \rangle \sigma_{\mu\nu}+\cdots \, ,
\end{eqnarray}

\begin{eqnarray}
S^{ij}_Q(x)&=&\frac{i}{(2\pi)^4}\int d^4k e^{-ik \cdot x} \left\{
\frac{\delta_{ij}}{\!\not\!{k}-m_Q}
-\frac{g_sG^n_{\alpha\beta}t^n_{ij}}{4}\frac{\sigma^{\alpha\beta}(\!\not\!{k}+m_Q)+(\!\not\!{k}+m_Q)
\sigma^{\alpha\beta}}{(k^2-m_Q^2)^2}\right.\nonumber\\
&&\left. -\frac{g_s^2 (t^at^b)_{ij} G^a_{\alpha\beta}G^b_{\mu\nu}(f^{\alpha\beta\mu\nu}+f^{\alpha\mu\beta\nu}+f^{\alpha\mu\nu\beta}) }{4(k^2-m_Q^2)^5}+\cdots\right\} \, , \end{eqnarray}
\begin{eqnarray}
f^{\lambda\alpha\beta}&=&(\!\not\!{k}+m_Q)\gamma^\lambda(\!\not\!{k}+m_Q)\gamma^\alpha(\!\not\!{k}+m_Q)\gamma^\beta(\!\not\!{k}+m_Q)\, ,\nonumber\\
f^{\alpha\beta\mu\nu}&=&(\!\not\!{k}+m_Q)\gamma^\alpha(\!\not\!{k}+m_Q)\gamma^\beta(\!\not\!{k}+m_Q)\gamma^\mu(\!\not\!{k}+m_Q)\gamma^\nu(\!\not\!{k}+m_Q)\, .
\end{eqnarray}
Then we compute  the integrals both in  coordinate space and in momentum space,  and obtain the correlation functions $\Pi_{J/\eta}(p^2)$ at the quark level, therefore the QCD spectral densities
 through dispersion relation.
 \begin{eqnarray}
{\lim}_{\epsilon\to 0} \frac{{\rm Im} \Pi_{J/\eta}(s+i\epsilon)}{\pi}&=&\rho_{J/\eta}(s)\, .
 \end{eqnarray}
 In Eqs.(14-15), we retain the terms $\langle\bar{q}_j\sigma_{\mu\nu}q_i \rangle$ and $\langle\bar{s}_j\sigma_{\mu\nu}s_i \rangle$  come  from the Fierz re-ordering   of the $\langle q_i \bar{q}_j\rangle$ and $\langle s_i \bar{s}_j\rangle$ to  absorb the gluons  emitted from other quark lines to form $\langle\bar{q}_j g_s G^a_{\alpha\beta} t^a_{mn}\sigma_{\mu\nu} q_i \rangle$ and $\langle\bar{s}_j g_s G^a_{\alpha\beta} t^a_{mn}\sigma_{\mu\nu} s_i \rangle$    to extract the mixed condensates  $\langle\bar{q}g_s\sigma G q\rangle$ and   $\langle\bar{s}g_s\sigma G s\rangle$, respectively.
 In this article, we carry out the
operator product expansion to the vacuum condensates  up to dimension-10, and take into account the vacuum condensates which are
vacuum expectations  of the operators  of the orders $\mathcal{O}( \alpha_s^{k})$ with $k\leq 1$ in a consistent way   \cite{WangHuang-3900,Wang-4660-2014,Wang-4025-CTP,WangHuang-NPA-2014,WangHuang-mole}.

Once the analytical expressions of the   QCD spectral densities $\rho_{J/\eta}(s)$ are obtained, we can  take the
quark-hadron duality below the continuum thresholds $s_0$ and perform Borel transform  with respect to
the variable $P^2=-p^2$ to obtain  the following QCD sum rules,
\begin{eqnarray}
\lambda^2_{Z}\, \exp\left(-\frac{M^2_{Z}}{T^2}\right)= \int_{4m_Q^2}^{s_0} ds\, \rho_{J/\eta}(s) \, \exp\left(-\frac{s}{T^2}\right) \, ,
\end{eqnarray}
where
\begin{eqnarray}
\rho_{J}(s)&=&\rho_{0}(s)+\rho_{3}(s)+\rho_{4}(s)+\rho_{5}(s)+\rho_{6}(s)+\rho_{8}(s)+\rho_{10}(s)\, , \\
\rho_\eta(s)&=&\rho_J(s)\mid_{m_s \to 0, \, \langle\bar{s}s\rangle \to \langle\bar{q}q\rangle,\,\langle\bar{s}g_s\sigma Gs\rangle \to \langle\bar{q}g_s\sigma Gq\rangle }\, ,
\end{eqnarray}

\begin{eqnarray}
\rho_0(s)&=&\frac{1}{512\pi^{6}}\int_{y_i}^{y_f}dy \int_{z_i}^{1-y}dz\,yz(1-y-z)^2\left(s-\overline{m}_Q^2\right)^3\left(5s-\overline{m}_Q^2\right) \nonumber\\
&&+\frac{m_Q^2}{128\pi^{6}}\int_{y_i}^{y_f}dy \int_{z_i}^{1-y}dz\,(1-y-z)^2\left(s-\overline{m}_Q^2\right)^3 \, ,
\end{eqnarray}

\begin{eqnarray}
\rho_3(s)&=&\frac{m_s\left[\langle\bar{s}s\rangle-2\langle\bar{q}q\rangle \right]}{32\pi^{4}}\int_{y_i}^{y_f}dy \int_{z_i}^{1-y}dz\,yz\left(s-\overline{m}_Q^2\right)\left(3s-\overline{m}_Q^2\right)  \nonumber\\
&&+\frac{m_s m_Q^2\left[\langle\bar{s}s\rangle-2\langle\bar{q}q\rangle \right]}{16\pi^{4}}\int_{y_i}^{y_f}dy \int_{z_i}^{1-y}dz\, \left(s-\overline{m}_Q^2\right) \, ,
\end{eqnarray}

\begin{eqnarray}
\rho_4(s)&=&-\frac{m_Q^2}{384\pi^{4}}\langle \frac{\alpha_sGG}{\pi}\rangle\int_{y_i}^{y_f}dy \int_{z_i}^{1-y}dz\,\left(\frac{z}{y^2}+\frac{y}{z^2} \right)(1-y-z)^2\left(2s-\overline{m}_Q^2\right)  \nonumber\\
&&-\frac{m_Q^4}{384\pi^{4}}\langle \frac{\alpha_sGG}{\pi}\rangle\int_{y_i}^{y_f}dy \int_{z_i}^{1-y}dz\,\left(\frac{1}{y^3}+\frac{1}{z^3} \right)(1-y-z)^2  \nonumber\\
&&+\frac{m_Q^2}{128\pi^{4}}\langle \frac{\alpha_sGG}{\pi}\rangle\int_{y_i}^{y_f}dy \int_{z_i}^{1-y}dz\,\left(\frac{1}{y^2}+\frac{1}{z^2} \right)(1-y-z)^2 \left(s-\overline{m}_Q^2\right)   \nonumber\\
&&-\frac{1}{1536\pi^{4}}\langle \frac{\alpha_sGG}{\pi}\rangle\int_{y_i}^{y_f}dy \int_{z_i}^{1-y}dz\,(1-y-z)^2\left(s-\overline{m}_Q^2\right)\left(5s-3\overline{m}_Q^2\right)    \nonumber\\
&&+\frac{1}{256\pi^{4}}\langle \frac{\alpha_sGG}{\pi}\rangle\int_{y_i}^{y_f}dy \int_{z_i}^{1-y}dz\,yz \left(s-\overline{m}_Q^2\right)\left(3s-\overline{m}_Q^2\right)    \nonumber\\
&&+\frac{m_Q^2}{128\pi^{4}}\langle \frac{\alpha_sGG}{\pi}\rangle\int_{y_i}^{y_f}dy \int_{z_i}^{1-y}dz\, \left(s-\overline{m}_Q^2\right)     \, ,
\end{eqnarray}

\begin{eqnarray}
\rho_5(s)&=& \frac{m_s\left[3\langle\bar{q}g_s\sigma Gq\rangle-\langle\bar{s}g_s\sigma Gs\rangle \right]}{48\pi^{4}}\int_{y_i}^{y_f}dy  \, y(1-y) s   \, ,
\end{eqnarray}

\begin{eqnarray}
\rho_6(s)&=&\frac{\langle\bar{q}q\rangle\langle\bar{s}s\rangle }{3\pi^{2}}\int_{y_i}^{y_f}dy \,y(1-y)s    \, ,
\end{eqnarray}

\begin{eqnarray}
\rho_8(s)&=&-\frac{\langle\bar{s}s\rangle\langle\bar{q}g_s\sigma Gq\rangle+\langle\bar{q}q\rangle\langle\bar{s}g_s\sigma Gs\rangle }{24\pi^{2}}\int_{y_i}^{y_f}dy \,y(1-y)\left[3+\left(4s+\frac{2s^2}{T^2}\right)\,\delta\left(s-\widetilde{m}_Q^2\right)\right]    \, , \nonumber\\
\end{eqnarray}

\begin{eqnarray}
\rho_{10}(s)&=& \frac{ \langle\bar{q}g_s\sigma Gq\rangle  \langle\bar{s}g_s\sigma Gs\rangle }{48\pi^{2}}\int_{y_i}^{y_f}dy \, y(1-y)\,\left(\frac{s}{T^2}+\frac{2s^2}{T^4}+\frac{s^3}{T^6} \right) \delta\left(s-\widetilde{m}_Q^2\right)  \nonumber\\
&&- \frac{ 11\langle\bar{q}g_s\sigma Gq\rangle  \langle\bar{s}g_s\sigma Gs\rangle }{6912\pi^{2}}\int_{y_i}^{y_f}dy \, \left(1+\frac{s}{2T^2} \right) \delta\left(s-\widetilde{m}_Q^2\right)  \, ,
\end{eqnarray}
  $y_{f}=\frac{1+\sqrt{1-4m_Q^2/s}}{2}$,
$y_{i}=\frac{1-\sqrt{1-4m_Q^2/s}}{2}$, $z_{i}=\frac{y
m_Q^2}{y s -m_Q^2}$, $\overline{m}_Q^2=\frac{(y+z)m_Q^2}{yz}$,
$ \widetilde{m}_Q^2=\frac{m_Q^2}{y(1-y)}$, $\int_{y_i}^{y_f}dy \to \int_{0}^{1}dy$, $\int_{z_i}^{1-y}dz \to \int_{0}^{1-y}dz$ when the $\delta$ functions $\delta\left(s-\overline{m}_Q^2\right)$ and $\delta\left(s-\widetilde{m}_Q^2\right)$ appear.

 We derive  Eq.(19) with respect to  $\tau=\frac{1}{T^2}$, then eliminate the
 pole residues   $\lambda_{Z}$ to obtain the QCD sum rules for the masses,
 \begin{eqnarray}
 M^2_{Z}= \frac{-\frac{d}{d \tau } \int_{4m_Q^2}^{s_0} ds\,\rho_{J/\eta}(s)\,e^{-\tau s}}{\int_{4m_Q^2}^{s_0} ds \,\rho_{J/\eta}(s)\,e^{-\tau s}}\, .
\end{eqnarray}

\section{Numerical results and discussions}

We take  the standard values of the vacuum condensates $\langle
\bar{q}q \rangle=-(0.24\pm 0.01\, \rm{GeV})^3$,   $\langle
\bar{q}g_s\sigma G q \rangle=m_0^2\langle \bar{q}q \rangle$,
$m_0^2=(0.8 \pm 0.1)\,\rm{GeV}^2$, $\langle\bar{s}s \rangle=(0.8\pm0.1)\langle\bar{q}q \rangle$, $\langle\bar{s}g_s\sigma G s \rangle=m_0^2\langle \bar{s}s \rangle$,  $\langle \frac{\alpha_s
GG}{\pi}\rangle=(0.33\,\rm{GeV})^4 $    at the energy scale  $\mu=1\, \rm{GeV}$
\cite{SVZ79,Reinders85,Colangelo-Review}, and choose the $\overline{MS}$ masses $m_{c}(m_c)=(1.275\pm0.025)\,\rm{GeV}$, $m_{b}(m_b)=(4.18\pm0.03)\,\rm{GeV}$,  $m_s(\mu=2\,\rm{GeV})=(0.095\pm0.005)\,\rm{GeV}$ from the Particle Data Group \cite{PDG}.
Furthermore, we take into account the energy-scale dependence of  the input parameters,
\begin{eqnarray}
\langle\bar{q}q \rangle(\mu)&=&\langle\bar{q}q \rangle(Q)\left[\frac{\alpha_{s}(Q)}{\alpha_{s}(\mu)}\right]^{\frac{4}{9}}\, , \nonumber\\
 \langle\bar{s}s \rangle(\mu)&=&\langle\bar{s}s \rangle(Q)\left[\frac{\alpha_{s}(Q)}{\alpha_{s}(\mu)}\right]^{\frac{4}{9}}\, , \nonumber\\
 \langle\bar{q}g_s \sigma Gq \rangle(\mu)&=&\langle\bar{q}g_s \sigma Gq \rangle(Q)\left[\frac{\alpha_{s}(Q)}{\alpha_{s}(\mu)}\right]^{\frac{2}{27}}\, , \nonumber\\ \langle\bar{s}g_s \sigma Gs \rangle(\mu)&=&\langle\bar{s}g_s \sigma Gs \rangle(Q)\left[\frac{\alpha_{s}(Q)}{\alpha_{s}(\mu)}\right]^{\frac{2}{27}}\, , \nonumber\\
m_c(\mu)&=&m_c(m_c)\left[\frac{\alpha_{s}(\mu)}{\alpha_{s}(m_c)}\right]^{\frac{12}{25}} \, ,\nonumber\\
m_b(\mu)&=&m_b(m_b)\left[\frac{\alpha_{s}(\mu)}{\alpha_{s}(m_b)}\right]^{\frac{12}{23}} \, ,\nonumber\\
m_s(\mu)&=&m_s({\rm 2GeV} )\left[\frac{\alpha_{s}(\mu)}{\alpha_{s}({\rm 2GeV})}\right]^{\frac{4}{9}} \, ,\nonumber\\
\alpha_s(\mu)&=&\frac{1}{b_0t}\left[1-\frac{b_1}{b_0^2}\frac{\log t}{t} +\frac{b_1^2(\log^2{t}-\log{t}-1)+b_0b_2}{b_0^4t^2}\right]\, ,
\end{eqnarray}
  where $t=\log \frac{\mu^2}{\Lambda^2}$, $b_0=\frac{33-2n_f}{12\pi}$, $b_1=\frac{153-19n_f}{24\pi^2}$, $b_2=\frac{2857-\frac{5033}{9}n_f+\frac{325}{27}n_f^2}{128\pi^3}$,  $\Lambda=213\,\rm{MeV}$, $296\,\rm{MeV}$  and  $339\,\rm{MeV}$ for the flavors  $n_f=5$, $4$ and $3$, respectively  \cite{PDG}, and evolve all the input parameters to the optimal energy scales   $\mu$ to extract the masses of the $Z_{QQ}$.

In Refs.\cite{WangHuang-3900,Wang-4660-2014,Wang-4025-CTP,WangHuang-NPA-2014,WangHuang-mole}, we study the acceptable energy scales of the QCD spectral densities  for the hidden-charm (hidden-bottom) tetraquark states and molecular states   in the QCD sum rules in details  for the first time,  and suggest an energy scale formula  $\mu=\sqrt{M^2_{X/Y/Z}-(2{\mathbb{M}}_Q)^2}$ to determine  the optimal   energy scales, which enhances the pole contributions remarkably and works well.
The  energy scale formula also works well in studying the hidden-charm pentaquark states \cite{WangPc}.    We can assign the $Z_c(3900)$ and $Z_b(10610)$ to be the axialvector tetraquark states with the quark constituents $cu\bar{c}\bar{d}$ and $bu\bar{b}\bar{d}$ respectively, and  choose the currents,
\begin{eqnarray}
J^{Q\bar{Q}}_\mu(x)&=&\frac{\varepsilon^{ijk}\varepsilon^{imn}}{\sqrt{2}}\left\{u^T_j(x)C\gamma_5Q_k(x) \bar{d}_m(x)\gamma_\mu C \bar{Q}^T_n(x)-u^T_j(x)C\gamma_\mu Q_k(x)\bar{d}_m(x)\gamma_5C \bar{Q}^T_n(x) \right\} \, , \nonumber\\
\end{eqnarray}
with $Q=c,\,b$ to study them with the QCD sum rules \cite{WangHuang-3900,WangHuang-NPA-2014}.  If we take the updated values of the effective heavy quark masses  ${\mathbb{M}}_c=1.82\,\rm{GeV}$ and ${\mathbb{M}}_b=5.17\,\rm{GeV}$ \cite{Wang-1601}, the optimal energy scales of the QCD spectral densities of the $Z_c(3900)$ and $Z_b(10610)$ are $\mu=1.4\,\rm{GeV}$ and $2.4\,\rm{GeV}$, respectively.

There are no experimental candidates for the  doubly heavy tetraquark states. Firstly, we suppose that the ground state $C\gamma_\mu\otimes \gamma_5C$ type axialvector tetraquark states $QQ\bar{u}\bar{d}$ and $Qu\bar{Q}\bar{d}$ have degenerate  masses, and study the masses of the ground state axialvector tetraquark states $QQ\bar{u}\bar{d}$ at the same energy scales of the QCD spectral densities as the ones for the ground state axialvector tetraquark states $Qu\bar{Q}\bar{d}$.
In Fig.1, we plot the predicted masses of the $Z_{cc\bar{u}\bar{d}}$ ($Z_{bb\bar{u}\bar{d}}$) and  $Z_c(3900)$ ($Z_b(10610)$)   with  variations  of the Borel parameter $T^2$ for the continuum threshold parameter $\sqrt{s_0}=4.4\,\rm{GeV}$ ($s_0^2=124\,\rm{GeV}^2$) and the energy scale $\mu=1.4\,\rm{GeV}$  ($\mu=2.4\,\rm{GeV}$) \cite{WangHuang-3900,WangHuang-NPA-2014,Wang-1601}. From the figure, we can see that the experimental values of the masses of the $Z_c(3900)$ and $Z_b(10610)$ can be well reproduced, there appear platforms for the masses of the $QQ\bar{u}\bar{d}$ tetraquark states, which lie slightly below the corresponding masses of the $Z_c(3900)$ and $Z_b(10610)$, respectively. If we choose the Borel windows as $T^2=(2.6-3.0)\,\rm{GeV}^2$ and $(6.9-7.7)\,\rm{GeV}^2$ for the tetraquark states $cc\bar{u}\bar{d}$ and $bb\bar{u}\bar{d}$, respectively, the pole contributions are $(44-58)\%$ and $(44-56)\%$, respectively, it is reliable to extract the masses. Furthermore, the continuum threshold parameters $s_0$ satisfy the relation $\sqrt{s_0}-M_{cc\bar{u}\bar{d}}=0.55\,\rm{GeV}$ and $\sqrt{s_0}-M_{bb\bar{u}\bar{d}}=0.62\,\rm{GeV}$, respectively, which are consistent with our naive expectation that the mass gaps of the ground states  and the first radial excited states of the tetraquark states are about $(0.5-0.6)\,\rm{GeV}$ \cite{Wang4430,Wang-3915-CgmCgm}. The energy scales $\mu=1.4\,\rm{GeV}$ and $2.4\,\rm{GeV}$ work well.

In Ref.\cite{KR-PRL},  Karliner and  Rosner obtain the masses $M=3.882\,\rm{GeV}$ and $10.389\,\rm{GeV}$  for the $C\gamma_\mu\otimes \gamma_5 C$ type axialvector tetraquark states  $cc\bar{u}\bar{d}$ and $bb\bar{u}\bar{d}$  respectively based on a simple potential quark model, which can reproduce the mass of the doubly charmed baryon state $\Xi^{++}_{cc}$.
In Ref.\cite{EQ-PRL}, Eichten and Quigg obtain the masses $M=3.978\,\rm{GeV}$ and $10.468\,\rm{GeV}$  for the $C\gamma_\mu\otimes \gamma_5 C$ type axialvector tetraquark states  $cc\bar{u}\bar{d}$ and $bb\bar{u}\bar{d}$   respectively based on the heavy quark symmetry, where the mass of the doubly charmed baryon state $\Xi^{++}_{cc}$ is taken as input parameter in the charm sector, while in the bottom sector, there are no experimental candidates for the baryon states $\Xi_{bb}^0$ and $\Xi_{bb}^-$.
From Fig.1, we can see that if we take the same parameters, such as the energy scales, continuum threshold parameters, etc, in the charm sector, the predicted mass $M_{cc\bar{u}\bar{d}}=3.85\,\rm{GeV}$ is slightly smaller than the value $3.882\,\rm{GeV}$ from  a simple potential quark model \cite{KR-PRL} and much smaller than the value $3.978\,\rm{GeV}$ from the heavy quark symmetry \cite{EQ-PRL},
in the bottom sector, the predicted mass  $M_{bb\bar{u}\bar{d}}=10.52\,\rm{GeV}$ is much larger than the value $10.389\,\rm{GeV}$ from  a simple potential quark model \cite{KR-PRL} and slightly larger than the value $10.468\,\rm{GeV}$ from the heavy quark symmetry \cite{EQ-PRL}.

\begin{figure}
 \centering
 \includegraphics[totalheight=5cm,width=7cm]{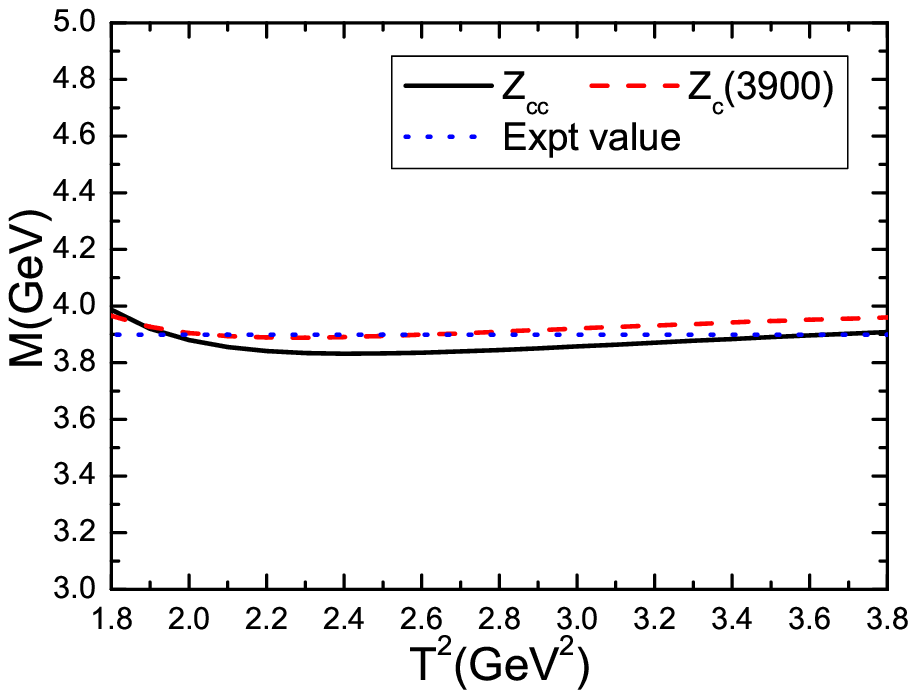}
 \includegraphics[totalheight=5cm,width=7cm]{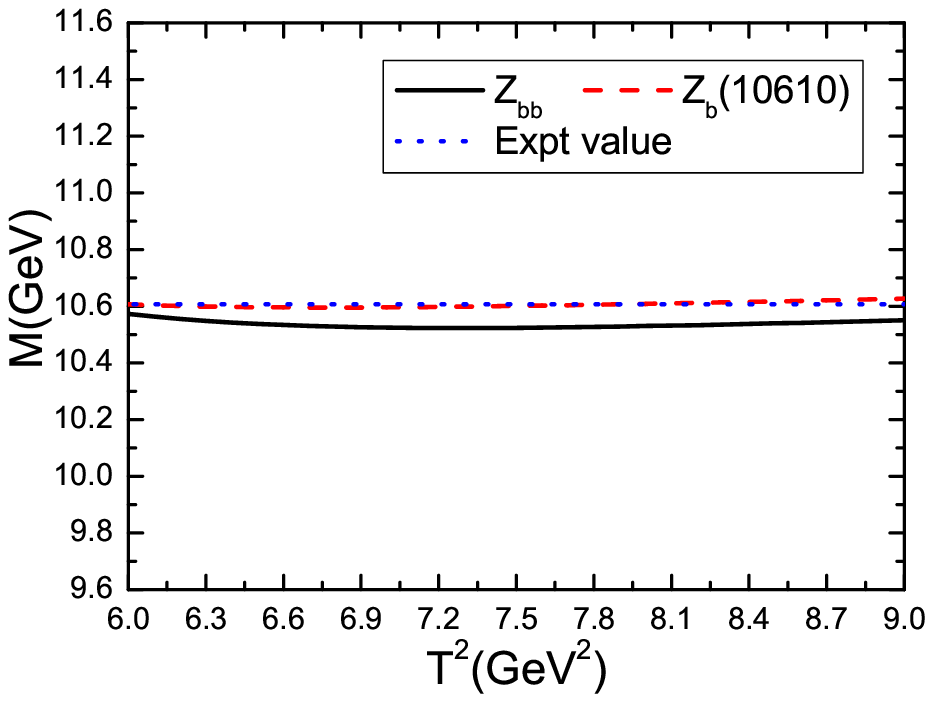}
        \caption{ The masses of the  $Z_{cc\bar{u}\bar{d}}$,  $Z_c(3900)$, $Z_{bb\bar{u}\bar{d}}$ and  $Z_b(10610)$    with variations  of the Borel parameter $T^2$, where the  Expt value denotes the experimental values of the masses $M_{Z_c(3900)}$ and $M_{Z_b(10610)}$.  }
\end{figure}

Now we revisit the subject of how to choose the energy scales of the QCD spectral densities.
In calculation, we  neglect  the perturbative
corrections to the currents $J/\eta_\alpha(x)$, which can be taken into
account in the leading logarithmic  approximation through  an anomalous dimension factor, $ \left[ \frac{\alpha_s(\mu_0)}{\alpha_s(\mu)}\right]^{\gamma_J}  $,
  the $\gamma_J$ are the anomalous dimension of the
 interpolating currents $J/\eta(x)$,
 \begin{eqnarray}
  \langle 0|J/\eta_\alpha(0;\mu)|Z_{QQ}(p)\rangle &=&\left[ \frac{\alpha_s(\mu_0)}{\alpha_s(\mu)}\right]^{\gamma_J}\langle 0|J/\eta_\alpha(0;\mu_{0})|Z_{QQ}(p)\rangle\nonumber\\
  &=&\left[ \frac{\alpha_s(\mu_0)}{\alpha_s(\mu)}\right]^{\gamma_J}\lambda_{Z}(\mu_0)\,\varepsilon_\alpha=\lambda_Z(\mu)\,\varepsilon_\alpha\, .
 \end{eqnarray}
The pole residues $\lambda_Z(\mu)=\left[ \frac{\alpha_s(\mu_0)}{\alpha_s(\mu)}\right]^{\gamma_J}\lambda_{Z}(\mu_0)$ are energy scale dependent quantities, at the
leading order approximation, we can set $\gamma_J=0$.

At the QCD side, the correlation functions $\Pi(p^2)$ can be written as
\begin{eqnarray}
\Pi(p^2)&=&\int_{4m^2_Q(\mu)}^{s_0} ds \frac{\rho_{J/\eta}(s,\mu)}{s-p^2}+\int_{s_0}^\infty ds \frac{\rho_{J/\eta}(s,\mu)}{s-p^2} \, ,
\end{eqnarray}
through dispersion relation, and they are energy scale independent according to the approximation $\gamma_J=0$ or $ \left[ \frac{\alpha_s(\mu_0)}{\alpha_s(\mu)}\right]^{\gamma_J}=1  $,
\begin{eqnarray}
\frac{d}{d\mu}\Pi(p^2)&=&0\, ,
\end{eqnarray}
which does not mean the pole contributions are energy scale independent,
\begin{eqnarray}
\frac{d}{d\mu}\int_{4m^2_Q(\mu)}^{s_0} ds \frac{\rho_{J/\eta}(s,\mu)}{s-p^2}\rightarrow 0 \, ,
\end{eqnarray}
 due to the following two reasons inherited from the QCD sum rules: (I) Perturbative corrections are neglected, the higher dimensional vacuum condensates are factorized into lower dimensional ones therefore  the energy scale dependence of the higher dimensional vacuum condensates is modified;
(II) Truncations $s_0$ set in, the correlation between the threshold $4m^2_Q(\mu)$ and continuum threshold $s_0$ is unknown,  the quark-hadron duality is just an assumption.
Even if the anomalous dimensions  $\gamma_J$ are neglected, the pole residues $\lambda_Z$ acquire energy scale dependence through the QCD side of the QCD sum rules, which does not mean that we cannot extract  reliable information of bound states.

In the article, we study the doubly heavy tetraquark states, the two heavy quarks form an axialvector  doubly heavy  diquark state in color antitriplet, then the axialvector doubly heavy  diquark state serves as a static well potential and combines with a light  antidiquark state in color triplet to form a compact tetraquark state. Such a tetraquark system is also characterized by the effective heavy quark mass ${\mathbb{M}}_Q$ and the virtuality $V=\sqrt{M^2_{X/Y/Z}-(2{\mathbb{M}}_Q)^2}$ (or bound energy not as robust) \cite{Wang-4660-2014,Wang-4025-CTP,WangHuang-NPA-2014}.   We obtain the energy scale formula  by setting the energy scale $\mu=V$.
It is not necessary  for the effective heavy quark masses ${\mathbb{M}}_Q$ in the doubly heavy tetraquark states to have the same values as the ones in the hidden-charm and hidden-bottom tetraquark states. In calculations, we observe that if we choose a slightly larger value ${\mathbb{M}}_c=1.84\,\rm{GeV}$, the criteria of the QCD sum rules (pole dominance at the hadron side and convergence of the operator product expansion at the QCD side) can be satisfied more easily, furthermore, other doubly charmed tetraquark states, such as the $C\gamma_\mu \otimes \gamma_\nu C$-type scalar, axialvector, tensor and vector tetraquark states,  can be described in the same routine \cite{Wang-Yan}. While in the bottom sector, a slightly smaller  value ${\mathbb{M}}_b=5.12\,\rm{GeV}$ does the work. In this article, we choose the values ${\mathbb{M}}_c=1.84\,\rm{GeV}$ and ${\mathbb{M}}_b=5.12\,\rm{GeV}$, and take into account the $SU(3)$ breaking effect $m_s(\mu)$ by subtracting the $m_s(\mu)$ from the  virtuality $V$, $\mu_k=V_k=\sqrt{M^2_{X/Y/Z}-(2{\mathbb{M}}_Q)^2}-k\,m_s(\mu_k)$, where the numbers of the strange antiquark $\bar{s}$ in the doubly heavy tetraquark states are $k=0,1$.
We cannot obtain energy scale independent QCD sum rules, but we have an energy scale formula  to determine the energy scales consistently.

\begin{table}
\begin{center}
\begin{tabular}{|c|c|c|c|c|c|c|c|}\hline\hline
                     &$T^2(\rm{GeV}^2)$   &$\sqrt{s_0}(\rm{GeV})$   &$\mu(\rm{GeV})$  &pole          &$M(\rm{GeV})$  &$\lambda(\rm{GeV}^5)$ \\ \hline

$cc\bar{u}\bar{d}$   &$2.6-3.0$           &$4.45\pm0.10$            &1.3              &$(39-63)\%$   &$3.90\pm0.09$  &$(2.64\pm0.42)\times10^{-2}$   \\ \hline
$cc\bar{u}\bar{s}$   &$2.6-3.0$           &$4.50\pm0.10$            &1.3              &$(41-64)\%$   &$3.95\pm0.08$  &$(2.88\pm0.46)\times10^{-2}$   \\ \hline

$bb\bar{u}\bar{d}$   &$6.9-7.7$           &$11.14\pm0.10$           &2.4              &$(41-60)\%$   &$10.52\pm0.08$ &$(1.30\pm0.20)\times10^{-1}$   \\ \hline
$bb\bar{u}\bar{s}$   &$6.8-7.6$           &$11.15\pm0.10$           &2.4              &$(41-61)\%$   &$10.55\pm0.08$ &$(1.33\pm0.20)\times10^{-1}$   \\ \hline

$cc\bar{u}\bar{d}$   &$2.6-3.0$           &$4.40\pm0.10$            &1.4              &$(39-62)\%$   &$3.85\pm0.09$  &$(2.60\pm0.42)\times10^{-2}$   \\ \hline
 \hline
\end{tabular}
\end{center}
\caption{ The Borel parameters (Borel windows), continuum threshold parameters, ideal energy scales, pole contributions,   masses and pole residues for the doubly heavy  tetraquark states. }
\end{table}

In this article, we take the continuum threshold parameters  as  $\sqrt{s_0}=M_{Z}+(0.4\sim0.7)\,\rm{GeV}$, and vary the parameters $\sqrt{s_0}$ to find  the optimal Borel parameters $T^2$ to satisfy the following four criteria:

$\bf 1.$ Pole dominance on the phenomenological side;

$\bf 2.$ Convergence of the operator product expansion;

$\bf 3.$ Appearance of the Borel platforms;

$\bf 4.$ Satisfying the energy scale formula.

The resulting  Borel parameters or Borel windows $T^2$, continuum threshold parameters $s_0$, optimal energy scales of the QCD spectral densities, pole contributions of the ground states are shown explicitly in Table 1. From Table 1, we can see that the pole dominance can be well satisfied. In Table 1, we also present the results where  the same parameters as the ones in the QCD sum rules for  the $Z_c(3900)$ are chosen, see the last line.

\begin{figure}
 \centering
 \includegraphics[totalheight=5cm,width=7cm]{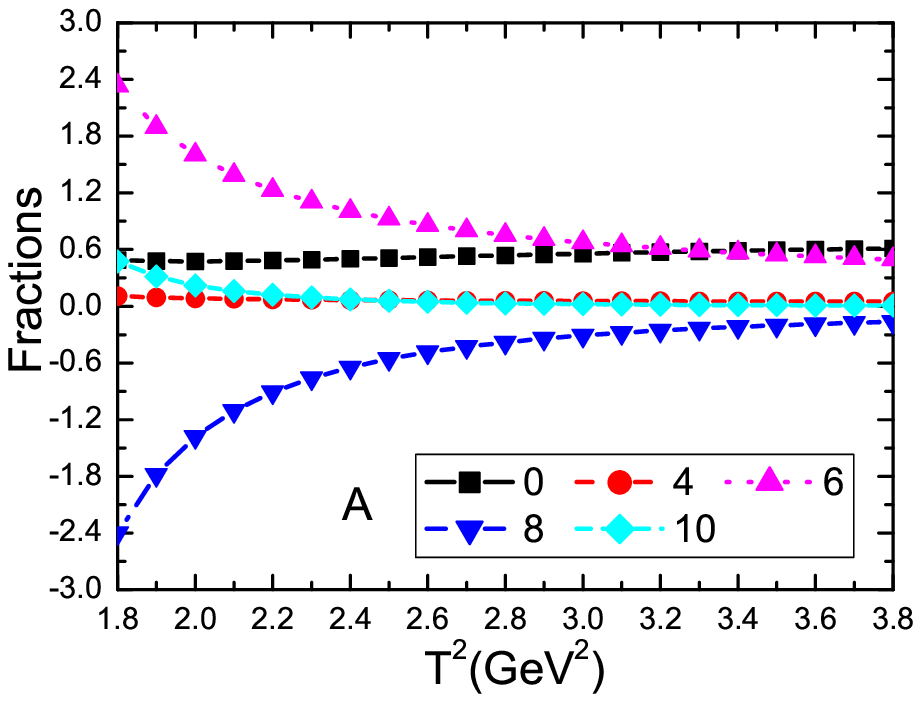}
  \includegraphics[totalheight=5cm,width=7cm]{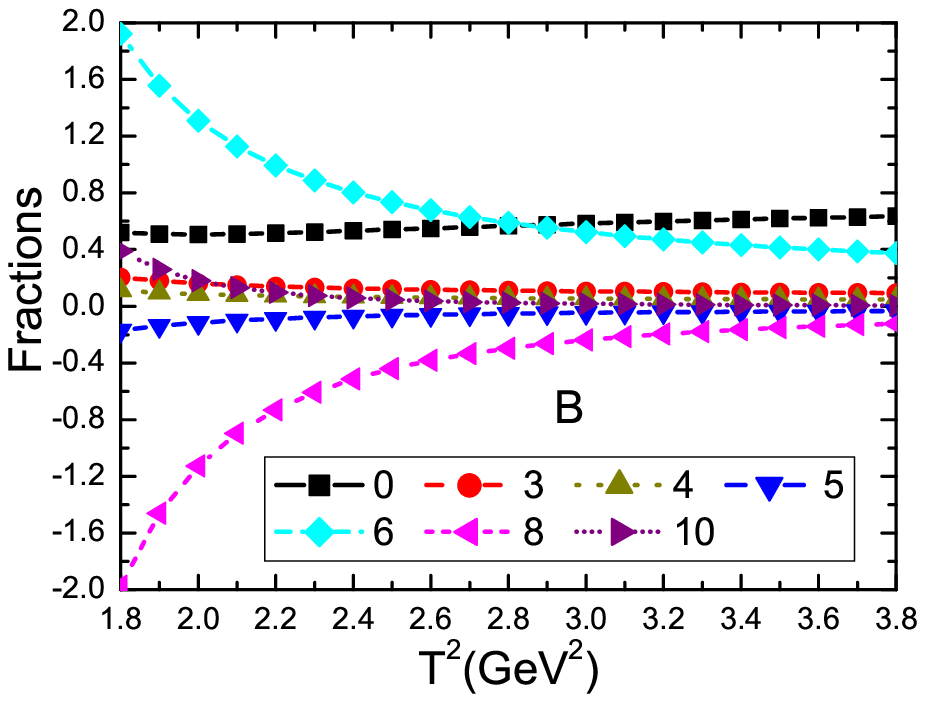}
  \includegraphics[totalheight=5cm,width=7cm]{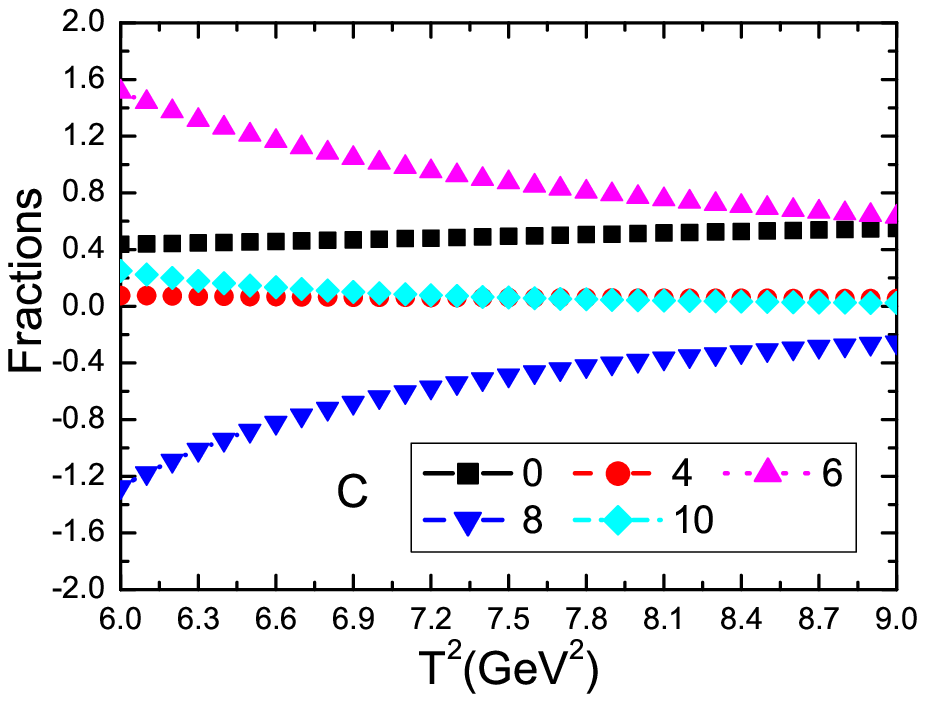}
  \includegraphics[totalheight=5cm,width=7cm]{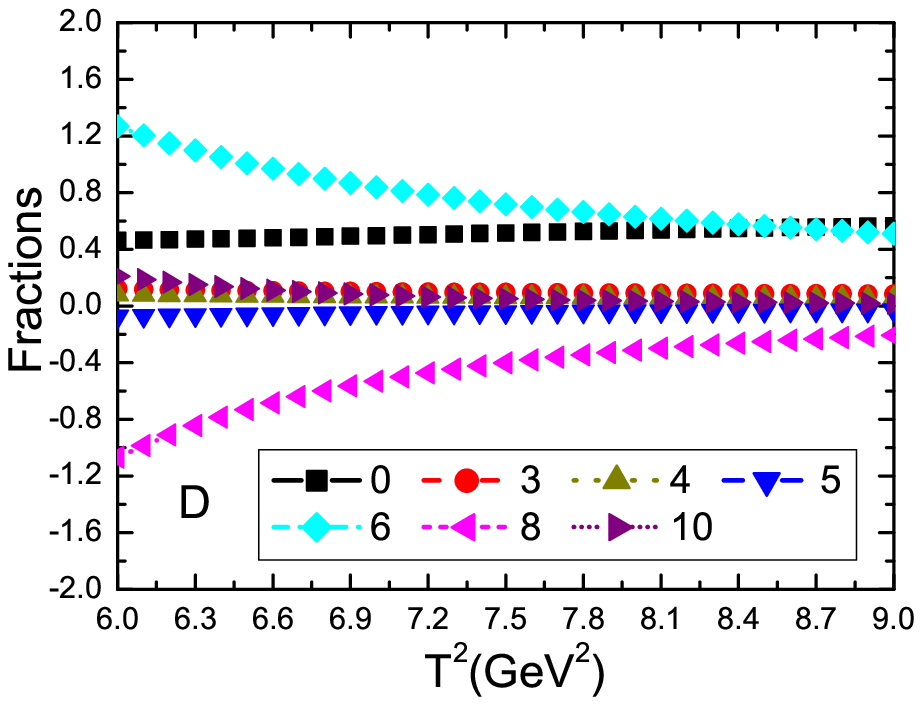}
        \caption{ The contributions of  different terms in the operator product expansion    with variations  of the Borel parameter $T^2$, where the $0$, $3$, $4$, $5$, $6$, $8$ and $10$ denote the dimensions of the vacuum condensates, the $A$, $B$, $C$ and $D$ denote the tetraquark states $cc\bar{u}\bar{d}$, $cc\bar{u}\bar{s}$, $bb\bar{u}\bar{d}$ and $bb\bar{u}\bar{s}$, respectively.  }
\end{figure}

 In Fig.2, we plot the contributions of the vacuum condensates in the operator product expansion with variations of the Borel parameter $T^2$ at much larger ranges than the Borel windows for the central values of the threshold parameters shown in Table 1.   From the figure, we can see that although the dominant contributions do not come from the perturbative terms,  the contributions of the vacuum condensates of dimensions $n=6$ and $8$ are very large, but the contributions of the vacuum condensates of dimensions $6, \,8,\,10$ have the hierarchy $D_6> |D_8|\gg D_{10}$ or $D_6\gg |D_8|\gg D_{10}$ in the Borel windows, the operator product expansion is convergent.

In the QCD sum rules for the tetraquark states and pentaquark states, the higher dimension vacuum condensates  are always
 factorized to lower dimension vacuum condensates with vacuum saturation,
  factorization works well in  large $N_c$ limit \cite{SVZ79}.  In reality, $N_c=3$, some  (not much) ambiguities maybe come from
the vacuum saturation assumption.  We choose  universal values for the   ${\mathbb{M}}_Q$, analogous pole contributions ($(40-60)\%$) and analogous criteria for the convergence of the operator product expansion ($D_6> |D_8|\gg D_{10}$ or $D_6\gg |D_8|\gg D_{10}$), the ambiguities are partially absorbed into the effective masses
${\mathbb{M}}_Q$. In previous works, we observed that
 vacuum saturation assumption works well for all the hidden-charm (hidden-bottom) tetraquark   (molecular) states and hidden-charm pentaquark states  \cite{WangHuang-3900,Wang-4660-2014,Wang-4025-CTP,WangHuang-NPA-2014,WangHuang-mole,WangPc}, the ambiguities originate from the  vacuum saturation cannot impair the predictive ability remarkably.

We take  into account all uncertainties of the input parameters,
and obtain the values of the masses and pole residues of
 the    $Z_{QQ}$, which are  shown explicitly in Table 1 and Figs.3-4.
From Figs.3-4, we can see that there appear platforms in  the Borel windows shown in Table 1.  Furthermore, from Table 1, we can see that the energy scale formula $\mu_k=\sqrt{M^2_{X/Y/Z}-(2{\mathbb{M}}_Q)^2}-k\,m_s(\mu_k)$ with $k=0,1$ is also satisfied. Moreover, from Table 1, we can see that the Borel parameters $T=(1.6-1.7)\,\rm{GeV}$ and $(2.6-2.8)\,\rm{GeV}$ for the doubly heavy tetraquark states $Z_{cc}$ and $Z_{bb}$, respectively, which satisfy the relation $\mu={\mathcal{O}}(T)$. In the regions $T=(1.6-1.7)\,\rm{GeV}$ and $(2.6-2.8)\,\rm{GeV}$ or $\mu=1.3\,\rm{GeV}$ and $2.4\,\rm{GeV}$,
we expect  to extract reliable information  of bound states.
Now the four criteria are all satisfied,  we expect to make reliable predictions.

\begin{figure}
 \centering
 \includegraphics[totalheight=5cm,width=7cm]{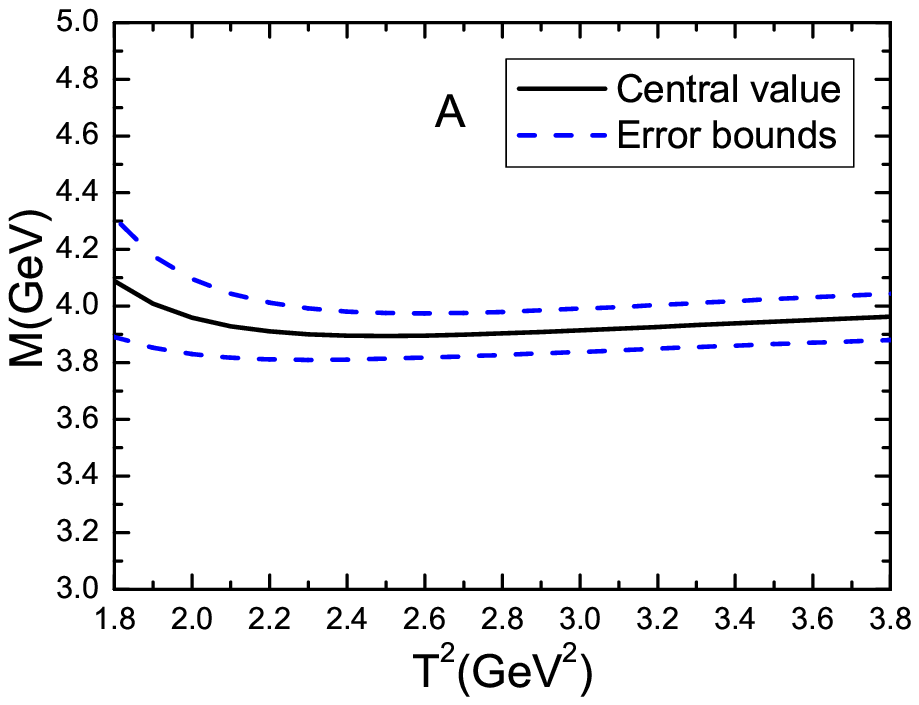}
  \includegraphics[totalheight=5cm,width=7cm]{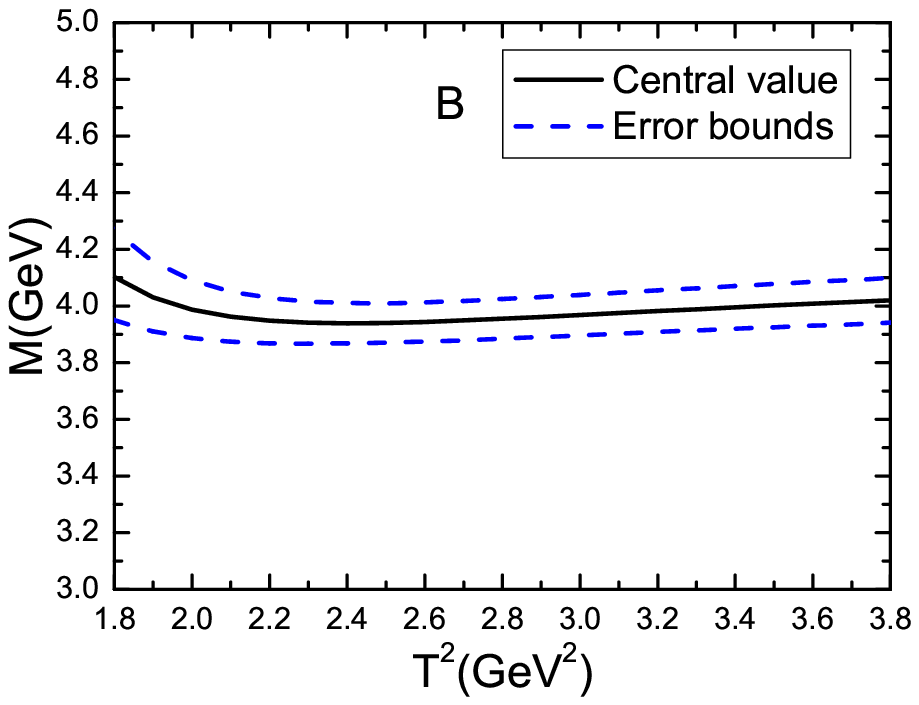}
  \includegraphics[totalheight=5cm,width=7cm]{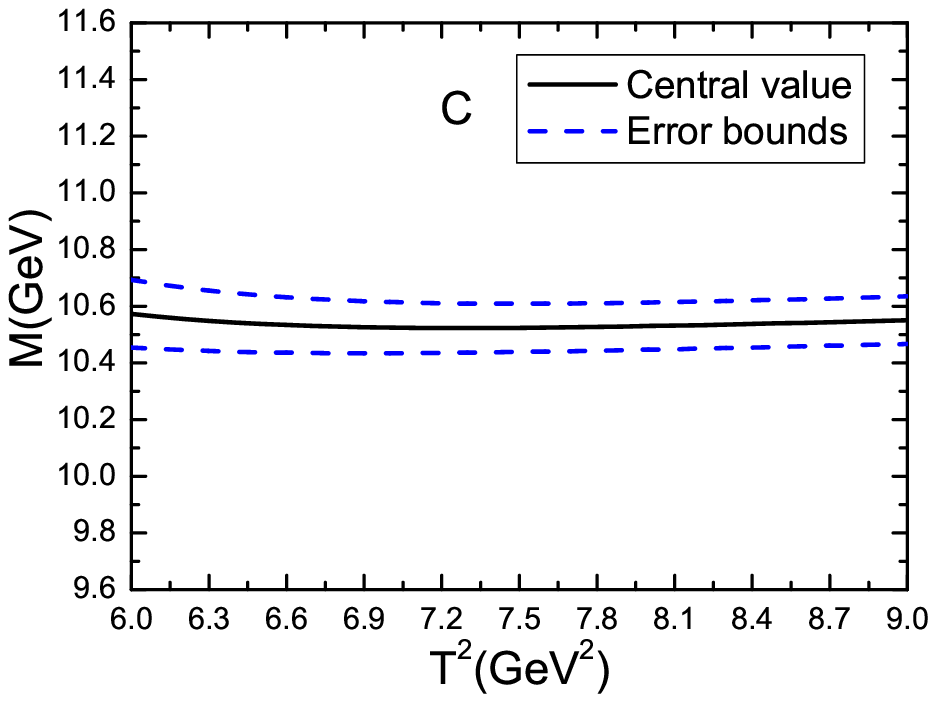}
  \includegraphics[totalheight=5cm,width=7cm]{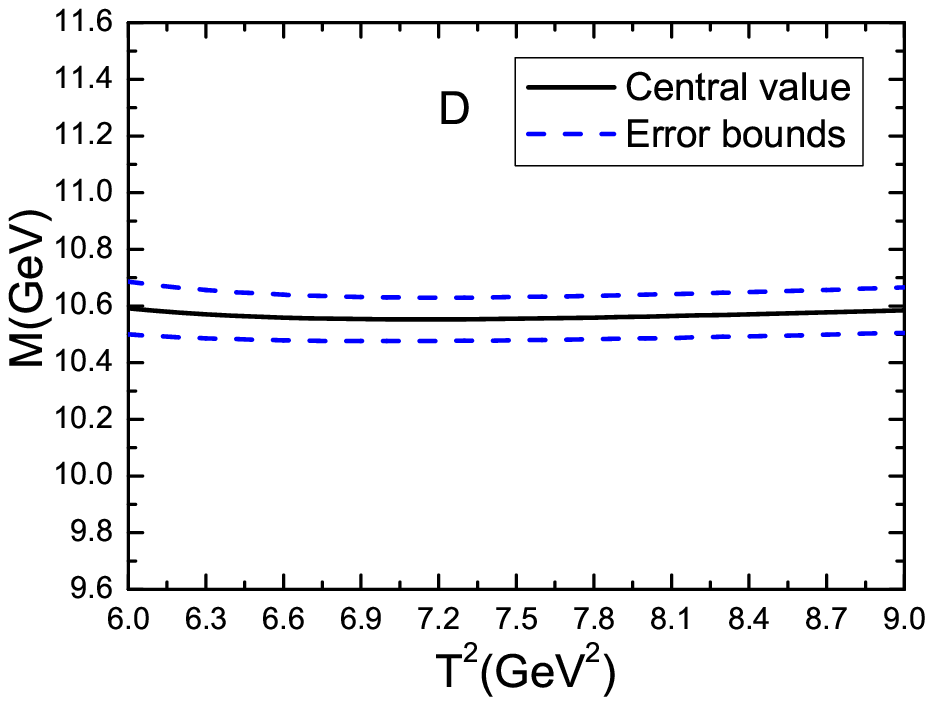}
        \caption{ The masses    with variations  of the Borel parameter $T^2$, where the  $A$, $B$, $C$ and $D$ denote the tetraquark states $cc\bar{u}\bar{d}$, $cc\bar{u}\bar{s}$, $bb\bar{u}\bar{d}$ and $bb\bar{u}\bar{s}$, respectively.  }
\end{figure}

\begin{figure}
 \centering
 \includegraphics[totalheight=5cm,width=7cm]{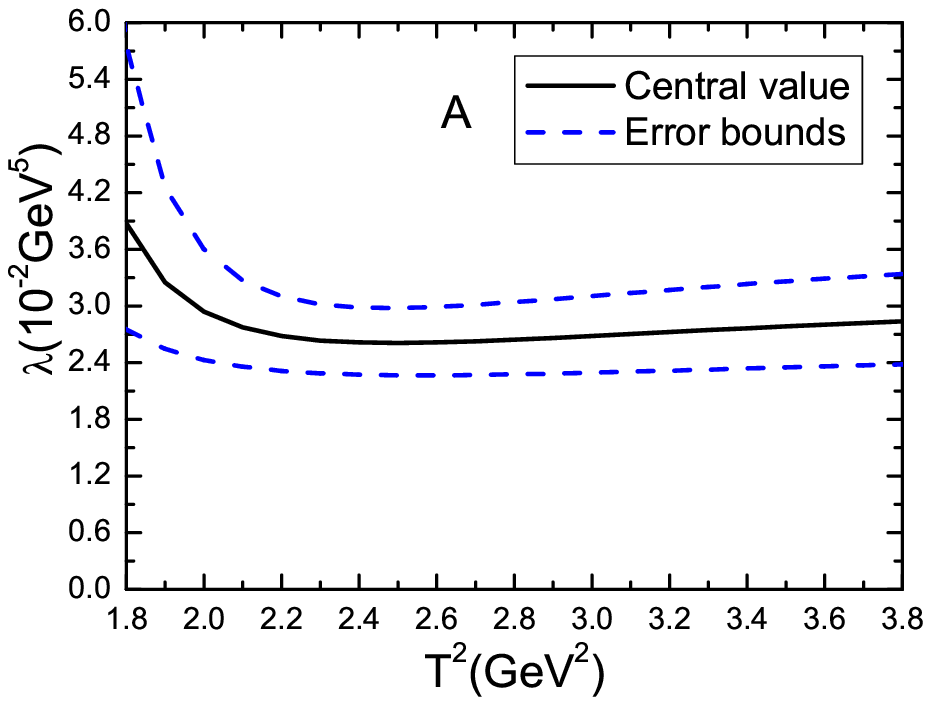}
  \includegraphics[totalheight=5cm,width=7cm]{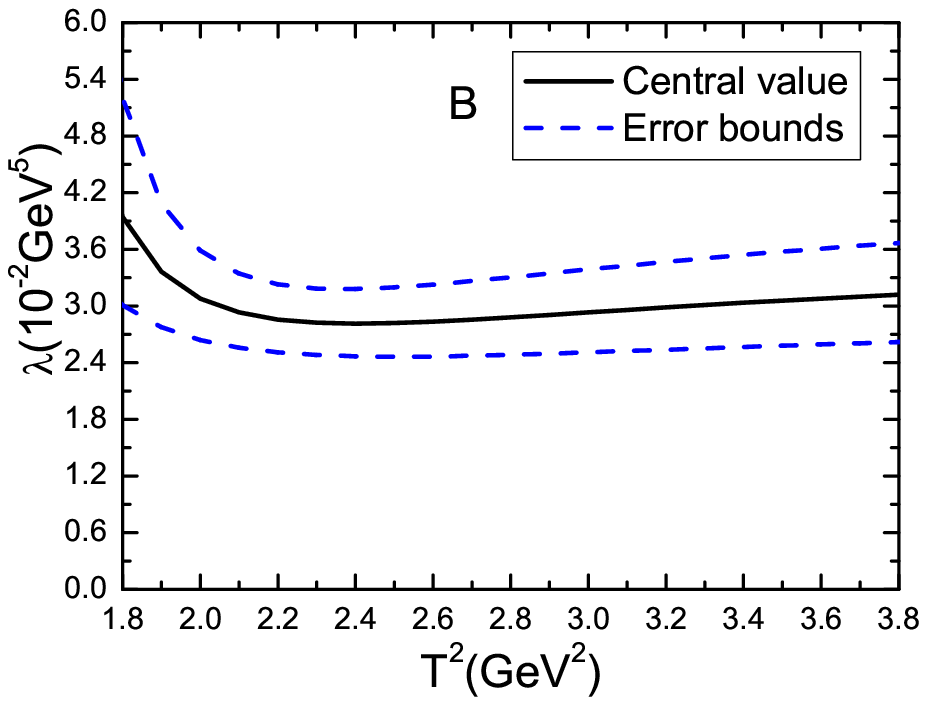}
  \includegraphics[totalheight=5cm,width=7cm]{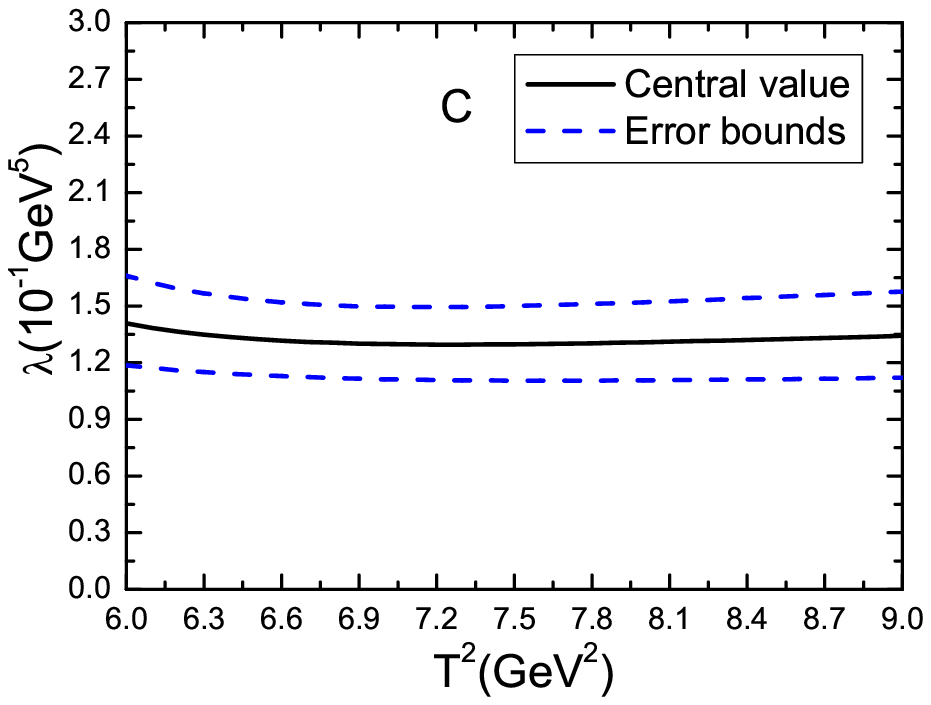}
  \includegraphics[totalheight=5cm,width=7cm]{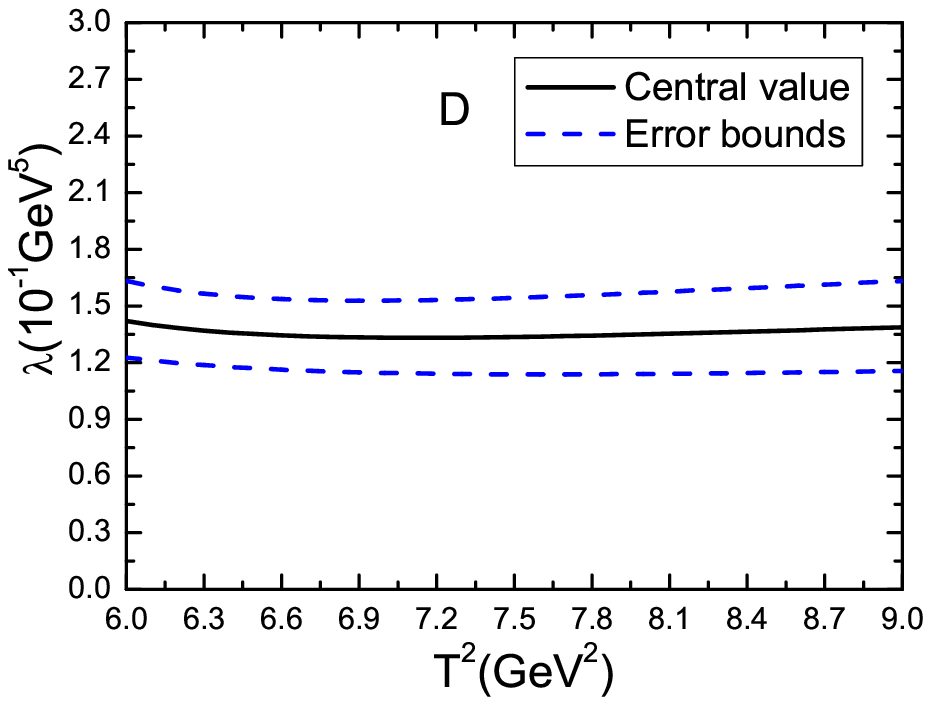}
        \caption{ The pole residues    with variations  of the Borel parameter $T^2$, where the  $A$, $B$, $C$ and $D$ denote the tetraquark states $cc\bar{u}\bar{d}$, $cc\bar{u}\bar{s}$, $bb\bar{u}\bar{d}$ and $bb\bar{u}\bar{s}$, respectively.  }
\end{figure}

\begin{figure}
 \centering
 \includegraphics[totalheight=5cm,width=7cm]{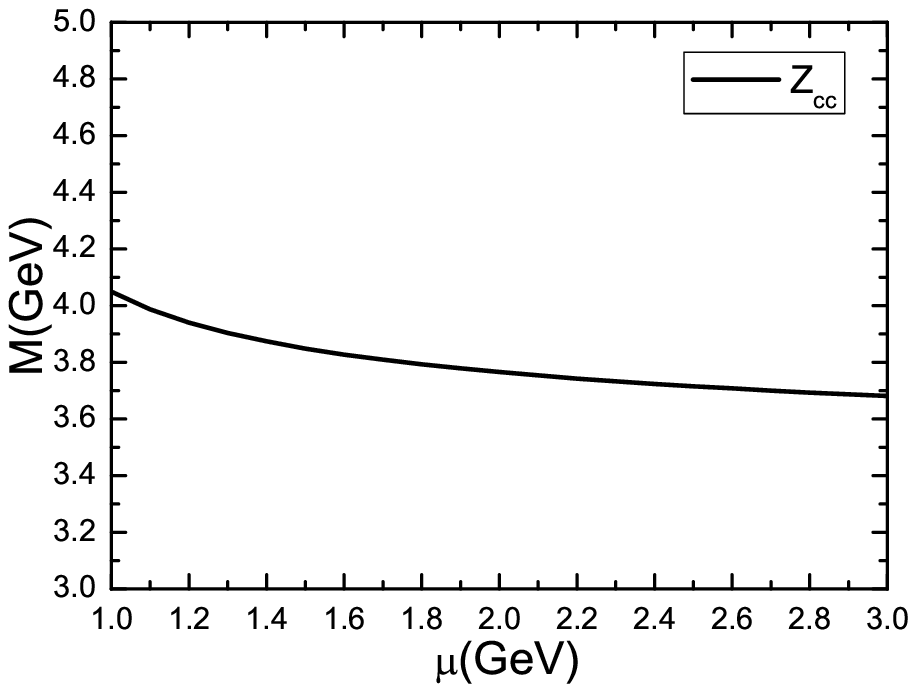}
 \includegraphics[totalheight=5cm,width=7cm]{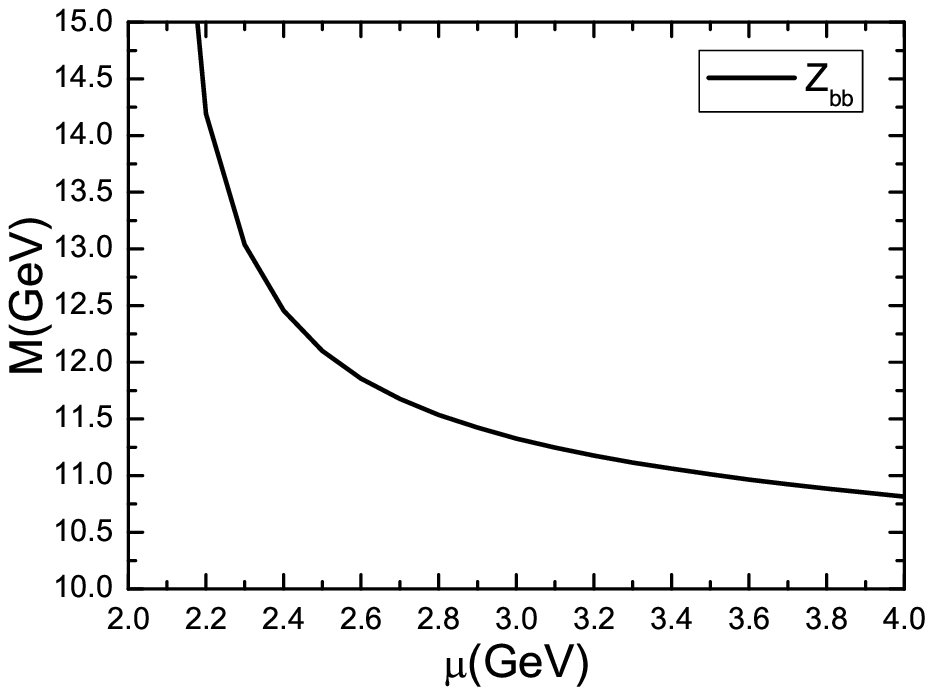}
        \caption{ The masses of the  $Z_{cc\bar{u}\bar{d}}$ and $Z_{bb\bar{u}\bar{d}}$     with variations  of the energy scale $\mu$.  }
\end{figure}

In this article, we have neglected  the perturbative $\mathcal{O}(\alpha_s)$ corrections to the perturbative terms. We can estimate the effects of the perturbative $\mathcal{O}(\alpha_s)$ corrections by multiplying the perturbative terms by a factor $1.3$ as the perturbative $\mathcal{O}(\alpha_s)$ corrections to the perturbative terms are usually about $30\%$. For example, we take into account the factor $1.3$ and refit the Borel window and threshold parameter for the $cc\bar{u}\bar{d}$ tetraquark state, and obtain the mass $3.91\,\rm{GeV}$ and pole residue $2.85\times 10^{-2}\,\rm{GeV}^5$, which is  consistent with the values  $M=3.90\,\rm{GeV}$ and $\lambda=2.64\times 10^{-2}\,\rm{GeV}^5$ in Table 1. So neglecting the perturbative $\mathcal{O}(\alpha_s)$ corrections cannot impair the predictive ability remarkably.

In Fig.5,  we plot the masses    with variations of the  energy scale $\mu$ for the central values of the input parameters shown in Table 1.
 From the figure, we can see that the masses decrease monotonously with increase of the energy scale, it is impossible to obtain energy scale independent QCD sum rules. In this article, we choose the special values determined by the energy scale formula in a consistent way.

\begin{table}
\begin{center}
\begin{tabular}{|c|c|c|c|c|c|c|c|}\hline\hline
                        &$cc\bar{u}\bar{d}$   &$cc\bar{u}\bar{s}$  &$bb\bar{u}\bar{d}$            &$bb\bar{u}\bar{s}$ \\ \hline

\cite{KR-PRL}           &$3.882 \pm 0.012$    &                    &$10.389\pm0.012$              &                    \\ \hline

\cite{EQ-PRL}           &$3.978$              &$4.156$             &$10.482$                      &$10.643$                    \\ \hline

\cite{QQ-Latt-mass1}    &                     &                    &$10.545^{+0.038}_{-0.030}$    &                    \\ \hline

\cite{QQ-Latt-mass2}    &                     &                    &$10.415 \pm0.0 10$            &$10.549 \pm0.008$                    \\ \hline

This work               &$3.90\pm0.09$        &$3.95\pm0.08$       &$10.52\pm0.08$                &$10.55\pm0.08$       \\ \hline

Thresholds              &$3.875/3.876$        &$3.977/3.976$       &$10.604/10.604$               &$10.692/10.695$       \\ \hline

 \hline
\end{tabular}
\end{center}
\caption{ The present predications compared to other theoretical works, where the Thresholds denote the two-meson thresholds $D^0D^{*+}/ D^+D^{*0}$, $D^0D_s^{*+}/D^+_sD^{*0}$, $\bar{B}^0B^{*-}/ B^-\bar{B}^{*0}$ and $\bar{B}_s^0B^{*-}/B^-\bar{B}_s^{*0}$, respectively, the unit is GeV. }
\end{table}

In Table 2, we list out the present predications compared to the values from some typical  theoretical approaches, such as the simple quark model \cite{KR-PRL}, heavy quark symmetry \cite{EQ-PRL}, lattice QCD \cite{QQ-Latt-mass1,QQ-Latt-mass2}. From the Table, we can see that the masses of the doubly charmed tetraquark states lie (slightly) above the corresponding lowest meson-meson thresholds, while the masses of the doubly bottom tetraquark states lie (slightly) below the corresponding lowest meson-meson thresholds, although the predicted masses differ from each other in one way or the other.

The decays of the doubly charmed (bottom) tetraquark states   $Z_{cc}$ ($Z_{bb}$) to the  charmed-meson (bottom meson) pairs are  Okubo-Zweig-Iizuka  super-allowed.
The two-body strong decays
\begin{eqnarray}
Z_{cc\bar{u}\bar{d}} &\to& D^0D^{*+}\, , \,\, D^+D^{*0}\, ,
\end{eqnarray}
are kinematically allowed, but the available phase spaces  are very small, if the hadronic coupling constant $G_{Z_{cc\bar{u}\bar{d}}DD^*}=G_{Z_c(3900)D\bar{D}^*}=0.62\,\rm{GeV}$ \cite{Wang-Zc3900-Decay}, then the width $\Gamma_{Z_{cc\bar{u}\bar{d}}}=0.44\,\rm{MeV}$. Even for large hadronic coupling constant $G_{Z_{cc\bar{u}\bar{d}}DD^*}=10\,G_{Z_c(3900)D\bar{D}^*}=6.2\,\rm{GeV}$, the width $\Gamma_{Z_{cc\bar{u}\bar{d}}}=44\,\rm{MeV}$ is still  negligible \cite{Wang-IJMPA}.
The two-body strong decays
\begin{eqnarray}
Z_{cc\bar{u}\bar{s}} &\to& D^0D_s^{*+}\, , \,\, D^+_sD^{*0}\, ,
\end{eqnarray}
can only take place for the upper bound of the predicted mass $M_{Z_{cc\bar{u}\bar{s}}}$, the width is expected to be  tiny.
While the two-body strong decays
 \begin{eqnarray}
Z_{bb\bar{u}\bar{d}} &\to& \bar{B}^0B^{*-}\, , \,\, B^-\bar{B}^{*0}\, , \nonumber\\
Z_{bb\bar{u}\bar{s}} &\to& \bar{B}_s^0B^{*-}\, , \,\, B^-\bar{B}_s^{*0}\, ,
\end{eqnarray}
 are kinematically forbidden, the $Z_{bb\bar{u}\bar{d}}$ and $Z_{bb\bar{u}\bar{s}}$ can decay weakly through $b\to c\bar{c}s$ at the quark level,
 \begin{eqnarray}
Z_{bb\bar{u}\bar{d}} &\to& \bar{B}^0B^{*-}\, , \,\, B^-\bar{B}^{*0}\to\, \gamma\, J/\psi K^-\,J/\psi \bar{K}^0 \,  , \nonumber\\
Z_{bb\bar{u}\bar{s}} &\to& \bar{B}_s^0B^{*-}\, , \,\, B^-\bar{B}_s^{*0}\to\, \gamma\, J/\psi \phi\, J/\psi K^-\,   ,
\end{eqnarray}
the widths can be neglected safely. The doubly charmed tetraquark states may be narrow resonances;  while the doubly bottom
 tetraquark states may be real ground tetraquark states and would establish the
existence of doubly bottom tetraquarks and illuminate the role of heavy
 diquarks in color antitriplet as the basic constituents.
According to the small or tiny widths of the lowest states, the one-pole approximation works well.
We can search for the doubly heavy tetraquark states in those decays in the future.

\section{Conclusion}
In this article, we construct the axialvector-diquark-scalar-antidiquark type currents to interpolate the axialvector doubly heavy tetraquark states, and study them with QCD sum rules by carrying out the operator product expansion up to the vacuum condensates of dimension 10. In calculations, we take the energy scale formula as a constraint to determine the energy scales of the QCD spectral densities in a consistent way to extract the masses and pole residues. In the Borel windows,  the pole dominance is satisfied  and  the operator product expansion is   convergent, and we expect to make reliable  predictions.  The present predictions indicate that the two body strong decays to the charmed   meson  pairs are  kinematically allowed, while two body strong decays to the bottom meson  pairs are  kinematically forbidden,  we can search for the axialvector doubly charmed (bottom) tetraquark states in strong (weak) decays in the future.

\section*{Acknowledgements}
This  work is supported by National Natural Science Foundation, Grant Number 11775079.

\end{document}